\documentclass[nofootinbib, pra, twocolumn]{revtex4}
\usepackage{amsmath}
\usepackage{amssymb}
\usepackage{graphicx}
\usepackage{bm}
\usepackage{braket}
\usepackage[version=4]{mhchem}
\usepackage{color, soul}
\usepackage[dvipsnames]{xcolor}
\usepackage{bbm}
\graphicspath{ {./} }
\usepackage{float}
\usepackage{soul}
\usepackage{capt-of}
\usepackage[colorlinks=true, citecolor=blue]{hyperref}

\newcommand{\beq}{\begin{equation}}
\newcommand{\eeq}{\end{equation}}

\definecolor{JM}{RGB}{4,116,149}

\definecolor{SJ}{RGB}{220,20,60}

\usepackage{listings}

\begin{document}

\title{Variational quantum algorithm for molecular geometry optimization}

\author{Alain Delgado}
\email{alaindelgado@xanadu.ai}
\affiliation{Xanadu, Toronto, ON, M5G 2C8, Canada}
\author{Juan Miguel Arrazola}
\affiliation{Xanadu, Toronto, ON, M5G 2C8, Canada}
\author{Soran Jahangiri}
\affiliation{Xanadu, Toronto, ON, M5G 2C8, Canada}
\author{Zeyue Niu}
\affiliation{Xanadu, Toronto, ON, M5G 2C8, Canada}
\author{Josh Izaac}
\affiliation{Xanadu, Toronto, ON, M5G 2C8, Canada}
\author{Chase Roberts}
\affiliation{Xanadu, Toronto, ON, M5G 2C8, Canada}
\author{Nathan Killoran}
\affiliation{Xanadu, Toronto, ON, M5G 2C8, Canada}

\begin{abstract}
Classical algorithms for predicting the equilibrium geometry of strongly correlated molecules
require expensive wave function methods that become impractical already for few-atom systems.
In this work, we introduce a variational quantum algorithm for finding the most stable structure of a molecule by explicitly considering the parametric dependence of the electronic Hamiltonian on the nuclear coordinates. The equilibrium geometry of the molecule is obtained by minimizing a more general cost function that depends on both the quantum circuit and the Hamiltonian parameters, which are simultaneously optimized at each step. The algorithm is applied to find the equilibrium geometries of the $\mathrm{H}_2$, $\mathrm{H}_3^+$, $\mathrm{BeH}_2$ and $\mathrm{H}_2\mathrm{O}$ molecules. The quantum circuits used to prepare the electronic ground state for each molecule were designed using an adaptive algorithm where excitation gates in the form of Givens rotations are selected according to the norm of their gradient. All quantum simulations are performed using the PennyLane library for quantum differentiable programming. The optimized geometrical parameters for the simulated molecules show an excellent agreement with their counterparts computed using classical quantum chemistry methods.
\end{abstract}

\maketitle

\section{Introduction}

In variational quantum algorithms for quantum chemistry, a quantum computer is programmed to prepare the wave function of a molecule and to measure the expectation value of the electronic Hamiltonian. A classical optimizer is then used to adjust the circuit parameters in order to minimize the total electronic energy \cite{peruzzo2014variational, mcardle2020quantum, cao2019quantum, cerezo2020variational}. Considerable attention has been placed on extending variational algorithms to compute excited-state energies \cite{higgott2019variational, mcclean2017hybrid, nakanishi2019subspace} and to mitigate the numerical errors inherent to noisy devices \cite{temme2017error, mcardle2019error}.

Extending the scope of quantum algorithms is crucial to study other molecular properties linked to the derivative of the total energy with respect to external parameters entering the electronic Hamiltonian~\cite{martinez2019mcvqe_gradients, mitarai2020gradients, azadsingh2021}. For example, computing the derivative of the energy with respect to the nuclear coordinates and external electric fields allows us to simulate the quantum vibrations of molecules and to predict their signature in experimental Raman and infrared spectra \cite{wilson1980molecular, pulay1987analytical}.

In particular, finding the equilibrium geometry of a molecule in a given electronic state is one of the most important tasks in computational quantum chemistry. Classical algorithms for molecular geometry optimization are computationally very expensive. They typically rely on the Newton-Raphson method requiring access to the nuclear gradients and the Hessian of the energy at each optimization step while searching for the global minimum along the potential energy surface \cite{jensen2017introduction}. As a consequence, using accurate post-Hartree-Fock methods \citep{jensen2017introduction} to solve the molecule's electronic structure at each step is computationally intractable even for medium-size molecules. Instead, density functional theory methods \cite{dft_book} are used to obtain approximated geometries.

In this work, we introduce a variational quantum algorithm for finding the equilibrium geometry of a molecule. We recast the problem as a more general variational quantum algorithm where the target electronic Hamiltonian is a parametrized observable that depends on the nuclear coordinates. This implies that the objective function, defined by the expectation value of the Hamiltonian computed in the trial state, depends on both the circuit and the Hamiltonian parameters. The proposed algorithm minimizes the cost function using a \emph{joint} optimization scheme where the analytical gradients of the cost function with respect to circuit parameters and the nuclear coordinates are computed simultaneously at each optimization step. Furthermore, this approach does not require nested optimizations of the circuit parameters for each set of nuclear coordinates, as occurs in the analogous classical algorithms. The optimized circuit parameters determine the energy of the electronic state prepared by the quantum circuit, and the final set of nuclear coordinates is precisely the equilibrium geometry of the molecule in this electronic state.

The manuscript is organized as follows. In Sec. \ref{sec:theory} we define the optimization problem and the methods to compute the quantum gradients of the cost function. Sec. \ref{sec:algorithm} describes each step of the quantum algorithm including its implementation using the PennyLane library for quantum differentiable programming \cite{bergholm2018pennylane}. In Sec. \ref{sec:application} we report numerical results on the geometry optimization of different molecules. The main conclusions are summarized in Sec. \ref{sec:conclusions}. 
  
\section{Theory}
\label{sec:theory}
We start by defining the parametrized Hamiltonian. For a molecule, this is the second-quantized electronic Hamiltonian for a given set of parameters $x$:

\begin{equation}
H(x) = \sum_{pq} h_{pq}(x)c_p^\dagger c_q + \frac{1}{2}\sum_{pqrs} h_{pqrs}(x) c_p^\dagger c_q^\dagger c_r c_s.
\label{eq:hamilt_x}
\end{equation}
The indices of summation in Eq.~\eqref{eq:hamilt_x} run over the basis of molecular orbitals computed in the Hartree-Fock approximation \cite{seeger1977self}. The operators $c^\dagger$ and $c$ are respectively the electron creation and annihilation operators, and $h_{pq}(x)$ and $h_{pqrs}(x)$ are the
one- and two-electron Coulomb integrals \cite{jensen2017introduction} computed in the molecular orbital basis.


In variational quantum algorithms, the expectation value of the target Hamiltonian is evaluated using a quantum computer, which is programmed to prepare a trial electronic wave function. To that aim, the Jordan-Wigner transformation \citep{seeley2012bravyi, jordan1928pauli} is typically applied to decompose the fermionic Hamiltonian in Eq.~\eqref{eq:hamilt_x} into a linear combination of Pauli operators,
\begin{equation}
H(x) = \sum_j h_j(x) \prod_i^{N} \sigma_i^j,
\label{eq:map_hamilt_x}
\end{equation}
where $h_j(x)$ are the expansion coefficients inheriting the dependence on the parameters $x$. The operators $\sigma_i$ represents the Pauli group $\{I, X, Y, Z\}$ and $N$ is the number of qubits

Let $\vert\Psi(\theta)\rangle$ denote the $N$-qubit trial state encoding the electronic state of the molecule that is implemented by a quantum circuit for a given set of parameters $\theta$. The expectation value of the parametrized Hamiltonian $H(x)$
\begin{equation}
g(\theta, x) = \langle \Psi(\theta) \vert H(x) \vert \Psi(\theta) \rangle,
\label{eq:cost_fn}
\end{equation}     
defines the cost function $g(\theta, x)$ for this problem, which can be optimized with respect to both the circuit and the Hamiltonian parameters. This is a generalization of the usual paradigm where only the state is parametrized. The variational quantum algorithm applied for solving the optimization problem
\begin{equation}
E = \min_{\{\theta, x\}} g(\theta, x),
\label{eq:optimization}
\end{equation} 
can be implemented to jointly optimize the circuit and Hamiltonian parameters $\theta$ and $x$, respectively. Crucially, the results of this optimization allow us to simultaneously find the lowest-energy state of the molecular Hamiltonian $\hat{H}(x)$ \emph{and} the optimal set of parameters $x$. For example, as we discuss later in this work, when the parameters correspond to the nuclear coordinates, the results of the optimization provide also the equilibrium geometry of the molecule.

Solving the optimization problem in Eq.~\eqref{eq:optimization} using gradient-based methods requires us to compute the gradients with respect to the circuit and the Hamiltonian parameters. The circuit gradients can be computed analytically using the parameter-shift rule \cite{schuld2019evaluating} in conjunction with the automatic differentiation algorithm, all of which are implemented in PennyLane \cite{bergholm2018pennylane}. The gradient with respect to the Hamiltonian parameters $x$ is obtained by evaluating the expectation value

\begin{equation}
\nabla_x g(\theta, x) = \langle \Psi(\theta) \vert \nabla_x H(x) \vert \Psi(\theta) \rangle.
\label{eq:gradient_x}
\end{equation}
The derivatives $\frac{\partial H(x)}{\partial x_i}$ of the Hamiltonian can be evaluated analytically or using finite differences. Analytical derivatives of the molecular Hamiltonian can be obtained in terms of the derivatives of the electron integrals $\frac{\partial h_{pq}(x)}{\partial x_i}$ and $\frac{\partial h_{pqrs}(x)}{\partial x_i}$. For example, if the parameters $x$ refer to the nuclear coordinates, the expressions to evaluate these derivatives have been established \cite{yamaguchi1994new} and they require solving the coupled-perturbed Hartree-Fock equations \cite{pulay1987analytical}.

\section{Quantum algorithm}
\label{sec:algorithm}
\begin{figure}[t]
\includegraphics[width=0.65 \columnwidth]{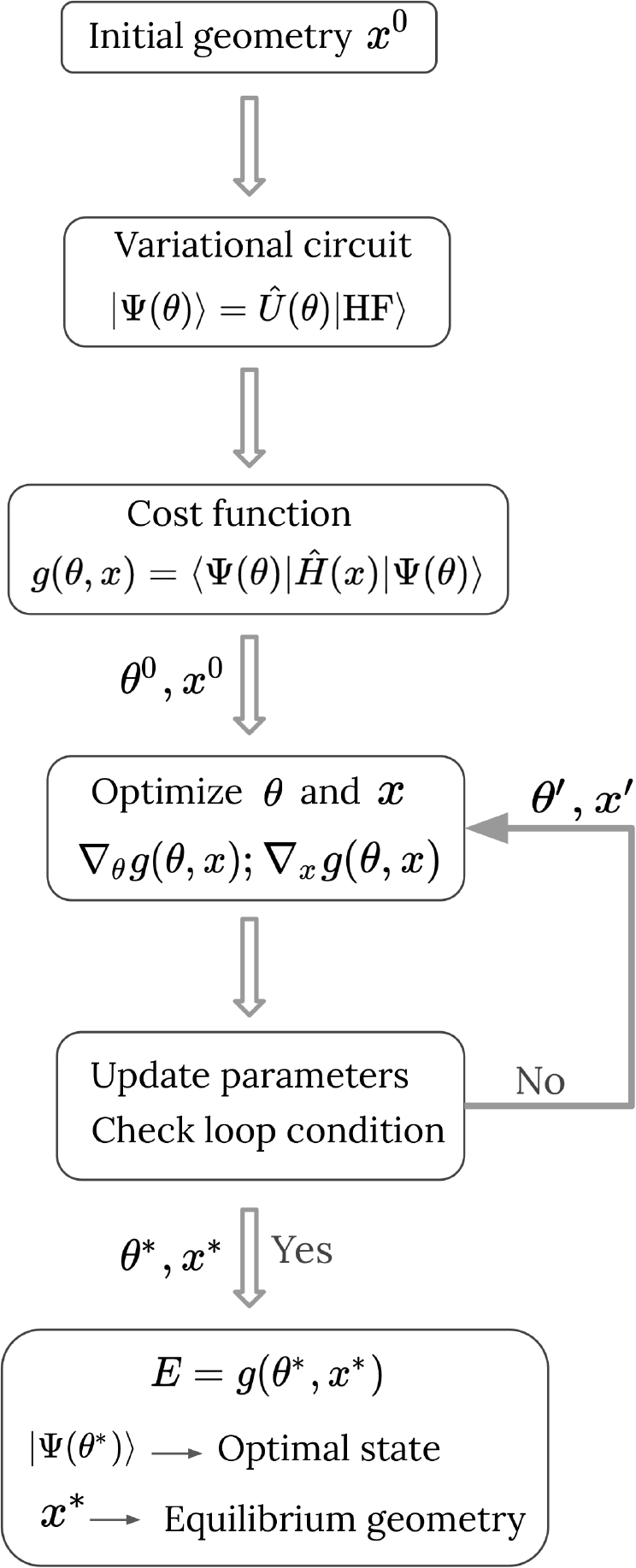}
\centering
\caption{Workflow of the variational quantum algorithm to find the equilibrium geometry of a molecule. The circuit parameters $\theta$ and the nuclear coordinates $x$ entering the Hamiltonian are jointly optimized. The iterative optimization is performed until a loop condition is satisfied. This could be a maximum number of iterations or a given convergence tolerance for the energy or the maximum component of the nuclear gradient.} \label{fig:algo_workflow}
\end{figure}

In this section we describe the quantum algorithm to solve the optimization problem of Eq.~\eqref{eq:optimization}. Without loss of generality, the algorithm is described for the problem of molecular geometry optimization where the Hamiltonian parameters $x$ are the nuclear coordinates of the molecule. The workflow of the algorithm is shown in Fig.~\ref{fig:algo_workflow}.
The algorithm takes as input the initial set of nuclear coordinates $x^0$ of the molecule we want to optimize. A good guess for the initial molecular geometry can be the geometry of the molecule optimized at the level of the Hartree-Fock (HF) approximation which can be efficiently computed using classical quantum chemistry packages.

We also need to define the variational quantum circuit to prepare the correlated electronic state $\vert\Psi(\theta)\rangle$ of the molecule. To that aim, the state of the $N$ qubits encoding the occupation number of the active spin-orbitals is initialized to encode the HF state. That is, the first $N_e$ qubits, with $N_e$ being the number of active electrons, are set in the state $\vert 1 \rangle$ while the other $N-N_e$ qubits remain in the state $\vert 0 \rangle$. The $N$-qubit system is then prepared in a superposition of the HF state with other doubly- and singly-excited configurations. In this work, this is done by applying \emph{excitation} gates implemented in the form of Givens rotations, as proposed in Ref.~[\onlinecite{paper_circuit_JM}]. Eq. \eqref{eq:se_givens} is an example of a Givens rotation:
\beq
G(\theta)=\begin{pmatrix}
1 & 0 & 0 & 0\\
0 & \cos (\theta) & -\sin (\theta) & 0\\
0 & \sin(\theta) & \cos(\theta) & 0\\
0 & 0 & 0 & 1
\end{pmatrix},\label{eq:se_givens}
\eeq
that acts as a \emph{single-excitation} two-qubit gate coupling the states $\vert 10 \rangle$ and $\vert 01 \rangle$ where a particle is ``excited'' from the first to the second qubit. Similarly, we also use the four-qubit \emph{double-excitation} gate $G^{(2)}$ to couple the states $\ket{1100}$ and $\ket{0011}$ differing by a double excitation,
\begin{align}
G^{(2)}\ket{1100} &= \mathrm{cos}(\theta)\ket{1100} - \mathrm{sin}(\theta)\ket{0011}, \\
G^{(2)}\ket{0011} &= \mathrm{cos}(\theta)\ket{0011} + \mathrm{sin}(\theta)\ket{1100}.
\end{align}
These excitation gates when applied to an $N$-qubit system act on the space of the specified qubits while acting as the identity on all other states \cite{paper_circuit_JM}.

Next, we can define the cost function $g(\theta, x)$ and proceed with the joint optimization of the circuit parameters $\theta$ and the nuclear coordinates $x$. The quantum gradients with respect to $\theta$ can be natively computed by PennyLane and the gradient with respect to $x$ is evaluated using Eq. (\ref{eq:gradient_x}). The derivative $\frac{\partial H(x)}{\partial x_i}$ of the electronic Hamiltonian is calculated using a central difference approximation. In our case, this is done by: i) displacing the $i$-th nuclear coordinate using a step of $0.01$ Bohr radii, ii) building the Hamiltonians corresponding to the perturbed coordinates, and iii) applying the finite-difference formula to build the observable $\frac{\partial H(x)}{\partial x_i}$ whose expectation value gives the $i$-th component of the nuclear gradient.

The cost function $g(\theta, x)$ is then minimized using a gradient-based optimizer until a maximum number of iterations is reached or a given convergence criterion is satisfied.
After the optimization is completed, the optimal parameters $\theta^*$ and $x^*$ can be used to compute the energy $E=g(\theta^*, x^*)$. The circuit parameters $\theta^*$ define the optimal electronic state of the molecule and the Hamiltonian parameters $x^*$ its equilibrium geometry in this electronic state.

As an example, the PennyLane code that implements step by step the quantum algorithm to optimize the geometry of the trihydrogen cation is given in the Appendix.

\section{Application: optimization of molecular geometries}
\label{sec:application}
\begin{figure}[h]
\includegraphics[width=1 \columnwidth]{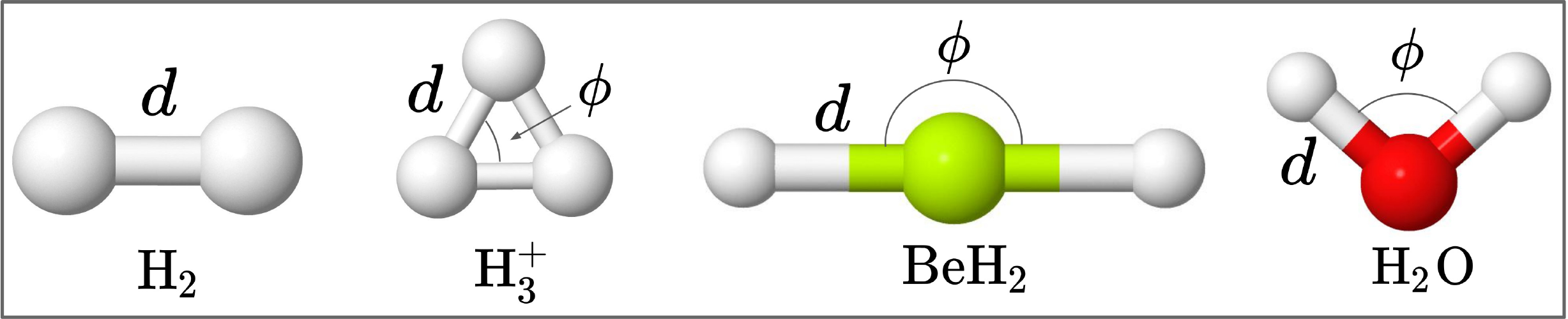}
\centering
\caption{Atomic structures and geometrical parameters of the simulated molecules. The parameters $d$ and $\phi$ denote the bond length and angle, respectively. The color code for the elements is white for hydrogen, green for beryllium, and red for oxygen.}\label{fig:mol_structures}
\end{figure}

We apply the quantum algorithm described in Sec. \ref{sec:algorithm} to find the ground-state equilibrium geometries of the hydrogen ($\mathrm{H_2}$), trihydrogen cation ($\mathrm{H_3}^+$), beryllium hydride ($\mathrm{BeH_2}$) and water ($\mathrm{H_2O}$) molecules. Their atomic structures are sketched in Fig.~\ref{fig:mol_structures}.

All the calculations were performed using the STO-3G minimal basis set \cite{basis_set_exchange}. In this approximation, for the $\mathrm{H}_2$ and $\mathrm{H}_3^+$ molecules we need four and six qubits, respectively, to encode the two-electron wave functions of these molecules. For the $\mathrm{BeH}_2$ and the $\mathrm{H_2O}$ molecules, the core electrons localized in the $s$-type orbitals of the beryllium and the oxygen atoms are excluded from the active space. That means that we have four and eight active electrons in the $\mathrm{BeH}_2$ and the $\mathrm{H_2O}$ molecules, respectively, whose wave functions are represented using twelve qubits.


We construct the variational circuit to prepare the correlated ground state for each of these molecules using an adaptive method similar to the algorithm proposed in Ref.~[\onlinecite{grimsley2019adaptive}]. We proceed as follows:

\begin{enumerate}
\item Generate all possible double excitations of the Hartree-Fock reference state. Typically, the dominant contributions to the ground-state correlation energy around the equilibrium geometry comes from the double excitations of the reference state.

\item Construct a circuit using all double-excitation gates acting on the qubits corresponding to the occupied and unoccupied orbitals involved in the double excitations.

\item Compute the gradient of the cost function with respect to each double-excitation gate and retain only those with non-zero gradient.

\item Optimize the parameters of the selected double-excitation gates.

\item Generate all single excitations of the Hartree-Fock state. Include the optimized double-excitation operations in the circuit. Apply all the single-excitation gates and select those with non-zero gradient.

\item Build the final variational quantum circuit by including the selected excitations.
\end{enumerate}

For the hydrogen molecule the variational circuit is very simple. We just apply a double-excitation gate to act on the four-qubit state $\ket{1100}$ to prepare the correlated state $\vert \Psi \rangle_{\mathrm{H}_2} = \mathrm{cos}(\theta) \ket{1100} - \mathrm{sin}(\theta) \ket{0011}$, where $\theta$ is the circuit parameter.
\begin{figure}[b]
\includegraphics[width=0.9\columnwidth]{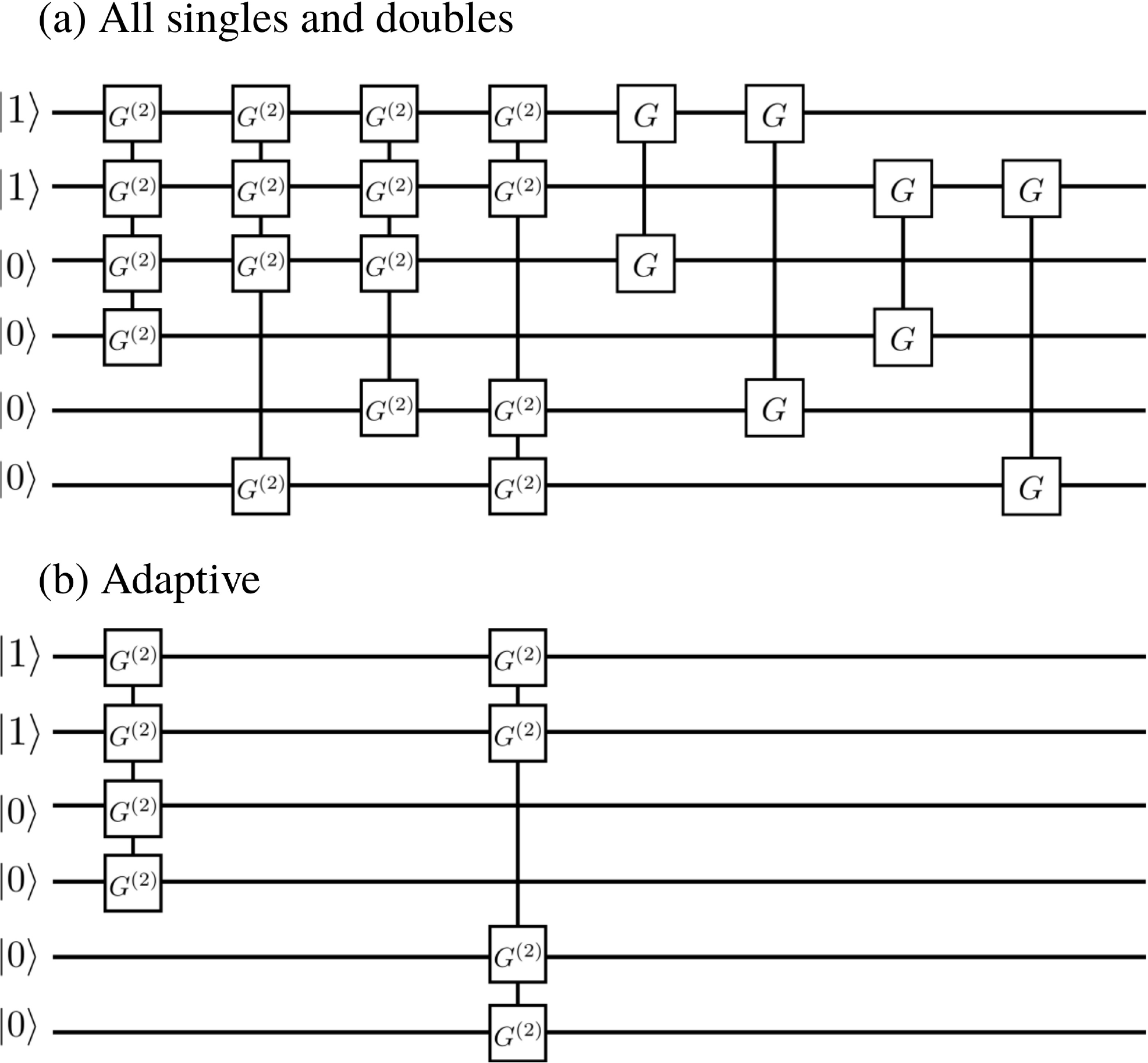}
\centering
\caption{Variational circuits to prepare the electronic ground state of the $\mathrm{H}_3^+$ molecule including (a) all single- and double-excitation operations preserving the total-spin projection of the HF state, (b) only the relevant excitation operations selected by the adaptive method. The squares indicate the qubits the operations act on.}
\label{fig:h3p_circuit}
\end{figure}
In the case of the $\mathrm{H}_3^+$ molecule, we have four singly- and four doubly-excited configurations that preserve the total-spin projection of the HF state. The quantum circuit including all corresponding single- and double-excitation operations is shown in Fig. \ref{fig:h3p_circuit}(a). By applying the adaptive method explained above the total number of gates is reduced from eight to two double-excitation gates, as shown in Fig. \ref{fig:h3p_circuit}(b). The selected gates are applied to prepare the trial state,
\begin{eqnarray}
\vert\Psi\rangle_{\mathrm{H}_3^+} &=& \mathrm{cos}(\theta_1)\mathrm{cos}(\theta_2)\vert110000\rangle - \mathrm{sin}(\theta_1)\vert001100\rangle \nonumber\\
&-& \mathrm{cos}(\theta_1)\mathrm{sin}(\theta_2)\vert000011\rangle.
\label{eq:state_h3p}
\end{eqnarray}

Similarly, we have built variational circuits for the beryllium hydride and the water molecules. As expected, the depth of the circuits increases with the system size. The total number of gates for these molecules, before and after applying the adaptive methodology, are reported in Table \ref{table:geo_results}.
\begin{figure}[b]
\includegraphics[width=0.9 \columnwidth]{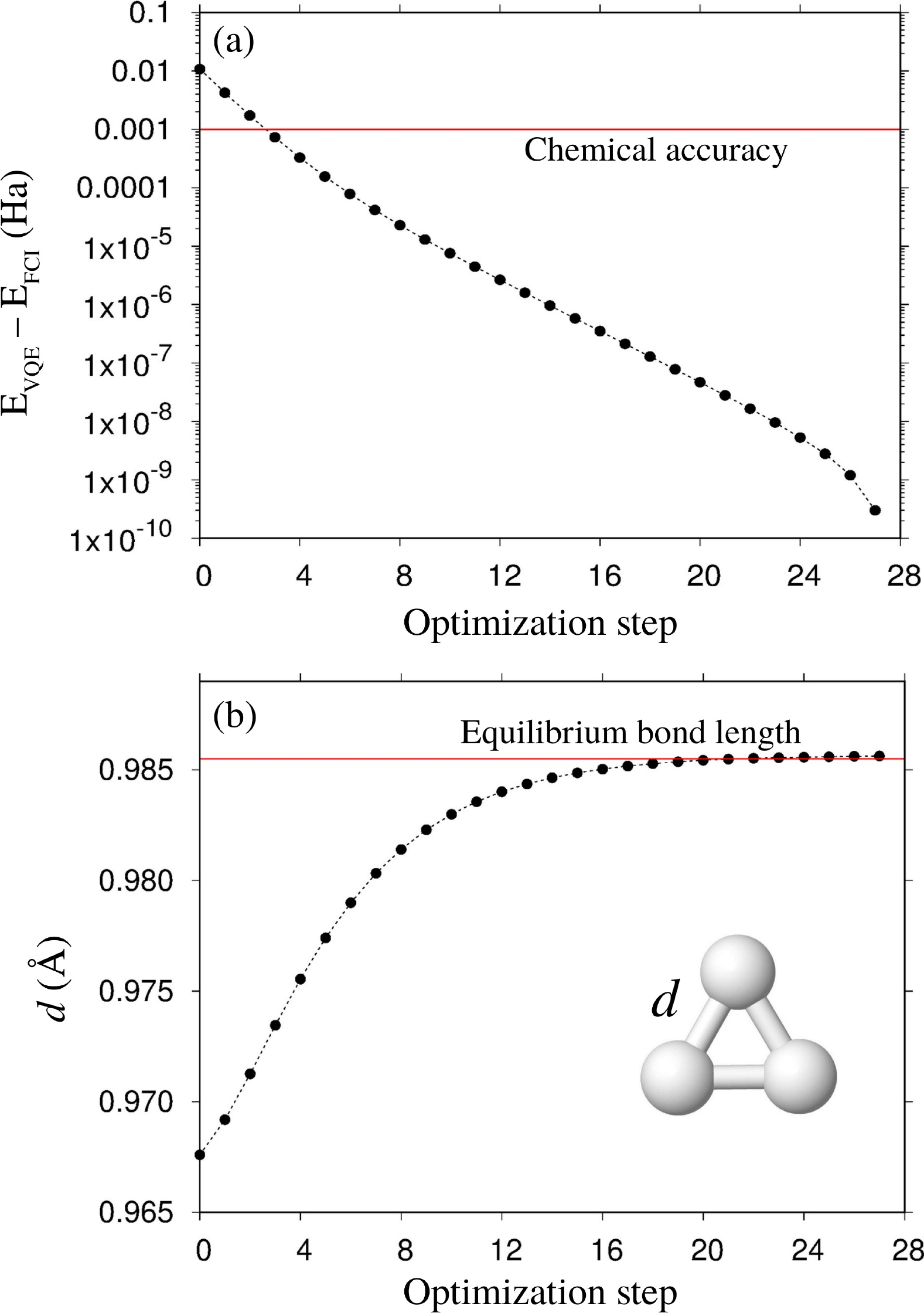}
\centering
\caption{(a) Convergence of the ground-state energy $E_\mathrm{VQE}$ and (b) the bond length $d$ for the $\mathrm{H}_3^+$ molecule as the circuit parameters and the nuclear coordinates are jointly optimized by the variational quantum algorithm. Values of the energy are reported relative to the analogous FCI value.}
\label{fig:opt_h3p}
\end{figure} 

We have performed numerical simulations of the quantum algorithm to find the equilibrium geometry of the molecules. All simulations were run using PennyLane. The nuclear coordinates were initialized to the Hartree-Fock geometry and the gate parameters were initially set to zero, i.e., we start from the Hartree-Fock state. The joint optimization of circuit parameters and nuclear coordinates was performed until the maximum component of the nuclear gradient was less than $10^{-5}$ Ha/Bohr.

In Fig.~\ref{fig:opt_h3p} we present numerical results for the geometry optimization of the $\mathrm{H}_3^+$ molecule. This figure plots the values of the ground-state energy of the molecule and the $\mathrm{H}$-$\mathrm{H}$ bond length as the circuit parameters $\theta$ and nuclear coordinates $x$ are jointly optimized by the quantum algorithm. The values of the energy are given relative to the exact value computed using the full configuration interaction (FCI) method as implemented in {\sc GAMESS} \cite{gamess_package}. From this figure we notice that despite the fact that the ground-state energy is already converged within chemical accuracy after the fourth step, more optimization steps are required to find the equilibrium bond length of the molecule. We observe in this case that an energy convergence tolerance of the order of $10^{-7}$ Ha would be required to find the optimized geometry of the molecule.

The results of the simulations are summarized in Table \ref{table:geo_results} where we report the geometrical parameters found by the quantum algorithm for all the simulated molecules. For comparison, the values of the bond lengths and angles obtained using the classical algorithms for molecular geometry optimization are given in parenthesis.
Remarkably, we observe an excellent agreement with these results for all the investigated molecules. We also note that the proposed adaptive method is useful to reduce the number of excitation gates in the variational circuits. In particular, for the $\mathrm{BeH}_2$ and $\mathrm{H_2O}$  molecules the total number of gates is significantly reduced from a total  of $92$ gates to $18$ and $30$ selected excitation gates, respectively.
\begin{table}[h]
\caption{Geometrical parameters of the optimized molecules. The values of the bond length $d$ and angle $\phi$ obtained with classical quantum chemistry calculations at the level of FCI are given in parenthesis. The number of qubits $N_\mathrm{qubits}$ corresponds to the number of active molecular spin-orbitals. $N_\mathrm{gates}$ is the number of excitation gates selected by the adaptive algorithm. For comparison, the total number of excitation gates before applying the adaptive method is given in parenthesis.}
\label{table:geo_results}
\begin{tabular}{c|c|c|c|c|c}
\hline\noalign{\smallskip}
Molecule & $N_e$ & $N_\mathrm{qubits}$ & $N_\mathrm{gates}$ & $d$ ($\mathrm{\AA}$)& $\phi$ (degrees) \\ \noalign{\smallskip}\hline\noalign{\smallskip}
 $\mathrm{H}_2$   & $2$ & $4$  & $1 ~ (1)$  &  $0.735 ~ (0.735)$   & -- \\
 $\mathrm{H}_3^+$ & $2$ & $6$  & $2 ~ (8)$  &  $0.986 ~ (0.986)$  & $60$ ($60$) \\
 $\mathrm{BeH}_2$ & $4$ & $12$ & $18 ~ (92)$  &  $1.316 ~ (1.316)$  & $180$ ($180$) \\
 $\mathrm{H_2O}$  & $8$ & $12$ & $30 ~ (92)$  &  $1.028 ~ (1.028)$  & $96.77$ ($96.74$) \\
\noalign{\smallskip}\hline\hline\noalign{\smallskip}
\end{tabular}
\end{table}

\section{Conclusions}
\label{sec:conclusions}
We have proposed a variational quantum algorithm to find the equilibrium geometry of molecules. We demonstrate that the stable structure of molecules can be found by minimizing a more general cost function that depends on the circuit parameters and the external parameters of the model Hamiltonian. Furthermore, we have shown that the minimization of the cost function can be achieved by jointly optimizing the circuit parameters defining the electronic state of the molecule and the nuclei positions. This joint optimization scheme does not require nested optimization of the circuit parameters as we update the nuclear coordinates of the molecule.

We used the variational quantum algorithm to find the equilibrium geometries of the $\mathrm{H}_2$, $\mathrm{H}_3^+$, $\mathrm{BeH}_2$ and the $\mathrm{H}_2\mathrm{O}$ molecules. We have used particle-conserving excitation gates to build the variational circuit preparing the electronic ground states of these molecules. We followed an adaptive algorithm to select quantum gates included in the variational circuit. The adaptive method has proven to be important to reduce the gate count and perform the molecular geometry optimizations. The use of a gradient-descent optimizer was found to be sufficient to converge to the equilibrium geometries of the investigated molecules. For all the simulated molecules we have found an excellent agreement of the optimized geometrical parameters with respect to the analogous results computed with traditional quantum chemistry simulations. 

\appendix

\section{PennyLane code implementing the proposed algorithm to optimize the geometry of the trihydrogen cation}
\label{code}

The python program implementing the quantum algorithm to optimize the ground-state geometry of the $\mathrm{H}_3^+$ molecule is shown below. This is a self-contained example illustrating how to implement the proposed variational quantum algorithm using the functionalities available in the PennyLane library.

The atomic species of the molecule are specified by the list \verb+symbols+ defined in line 6. The function \verb+H(x)+ builds the qubit Hamiltonian of the $\mathrm{H}_3^+$ molecule for a given set of the nuclear coordinates \verb+x+ using the \verb+molecular_hamiltonian()+ function.

The function \verb+circuit+ in line 13 uses PennyLane quantum operations to define the variational circuit shown in Fig. \ref{fig:h3p_circuit}(b) preparing the ground state of the $\mathrm{H}_3^+$ molecule. First, the \verb+qml.BasisState+ operation initializes the qubit register to the Hartree-Fock state. Then, two double-excitation gates acting on the qubits $[0, 1, 2, 3]$ and $[0, 1, 4, 5]$ are applied using the \verb+qml.DoubleExcitation+ operation. The circuit implemented by this function prepares the correlated state defined in Eq. \eqref{eq:state_h3p} that we use to compute the expectation values of the fermionic observables.

Line 20 declares the device \verb+dev+ on which we run the quantum algorithm. In this case we use the \verb+default.qubit+ simulator with a total of six qubits (\verb+wires+).

The objective function $g(\theta, x)$ defined in Eq. \eqref{eq:cost_fn} is implemented by the function \verb+cost(x, params)+ in line 23. It returns the expectation value of the parametrized Hamiltonian \verb+H(x)+ computed in the trial state prepared by the \verb+circuit+ function for a given set of parameters \verb+params+. The expectation value is calculated using the \verb+ExpvalCost+ function.

In line 27 we define the function \verb+grad_x(x, params)+ to compute the gradient of the cost function $g(\theta, x)$ with respect to the nuclear coordinates following Eq. \eqref{eq:gradient_x}. The nuclear gradient of the electronic Hamiltonian \verb+H(x)+ is calculated using the finite-difference function \verb+qml.finite_diff+.

We define the classical optimizers using the \verb+qml.GradientDescentOptimizer+ function and set the starting geometry of the molecule and the initial values of the circuit parameters in lines 37 and 42, respectively. Finally, the equilibrium geometry of the molecule is found by performing an iterative optimization of the circuit parameters \verb+theta+ and the nuclear coordinates \verb+x+ until the maximum component of the nuclear gradient is less than or equal to $10^{-5}$ Ha/Bohr.

\begin{figure*}[t]
\includegraphics[width=1.5 \columnwidth]{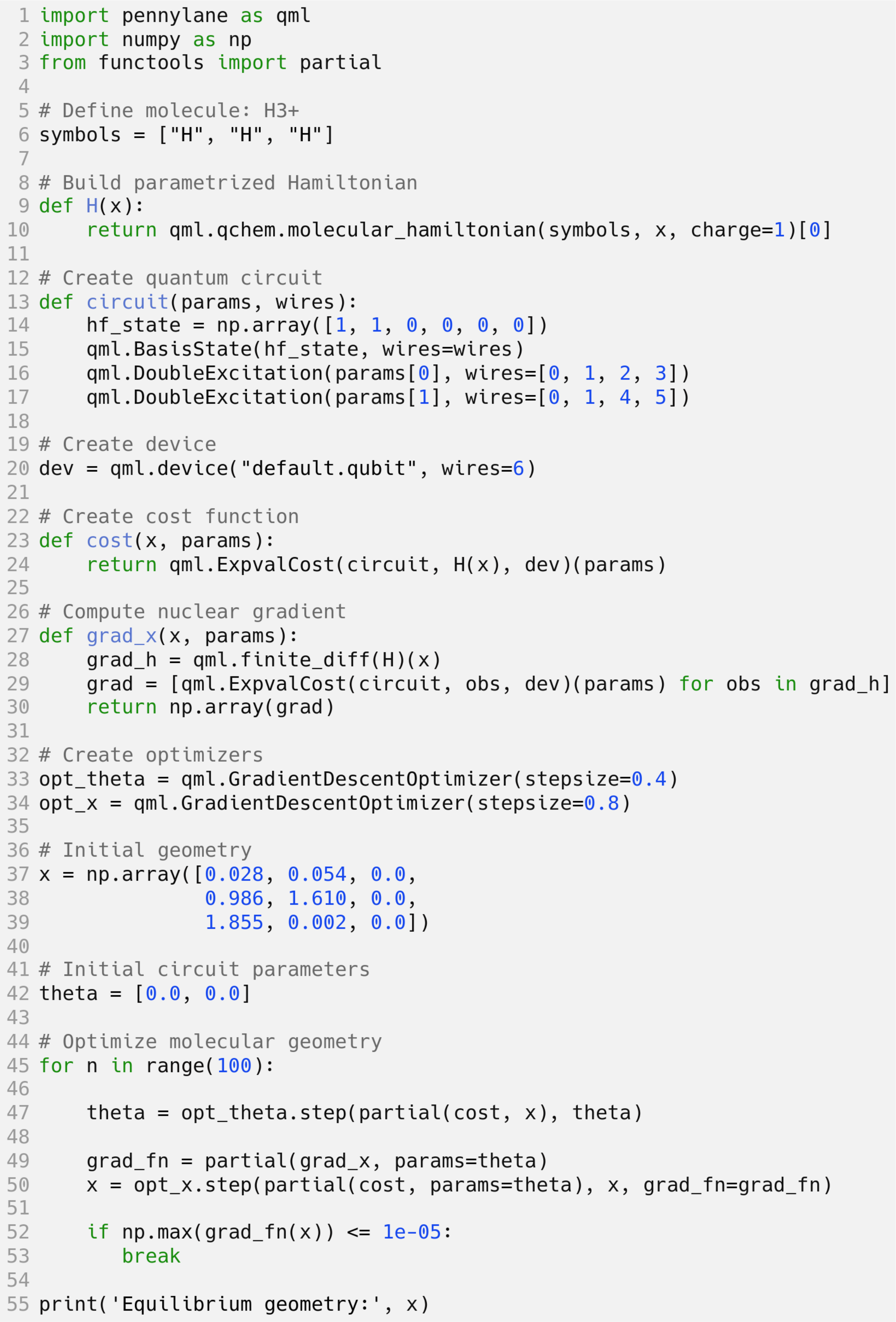} \\
\label{fig_code}
\begin{flushleft}
Code implementing the quantum algorithm to optimize the ground-state geometry of the $\mathrm{H}_3^+$ molecule using the PennyLane library.
\end{flushleft}
\end{figure*}

\newpage
\bibliographystyle{apsrev}
\bibliography{geo_opt}

\end{document}